%% file: main.tex
 \documentclass[smallabstract,smallcaptions]{dccpaper}

\usepackage{amsmath,stmaryrd}
\usepackage{color}
\usepackage{dsfont}
\usepackage{url}
\usepackage{graphicx}
\usepackage{subfigure}

\usepackage{caption}
\usepackage{lipsum}
\usepackage{algorithmic}

 \usepackage{amssymb}
\usepackage{soul}
\usepackage{mathrsfs}

\usepackage{setspace}

\usepackage{mathtools}

\input{macros}

\newlength{\figurewidth}
\newlength{\smallfigurewidth}

\begin{document}

\title
{\large
\textbf{Rate-Distortion Performance of Sequential Massive Random Access to Gaussian Sources with Memory}
}

\author{Elsa Dupraz$^{1}$, Thomas Maugey$^2$, Aline Roumy$^2$, Michel Kieffer$^3$ \\ ~ \\ 
\small $^1$ IMT Atlantique, Lab-STICC, UBL, \small $^2$ INRIA Rennes Bretagne-Atlantique, \\ \small $^3$ L2S, UMR CNRS 8506; CentraleSupelec; Univ. Paris-Sud}

%

\maketitle
\thispagestyle{empty}

\input{0abstract}

\input{1intro}

\input{2definition}
\input{3stateofart}
\input{4results}
\input{5conclusion}

\Section{References}
\bibliographystyle{ieeetr}
\bibliography{biblio}

\end{document}

%% file: macros.tex
\usepackage{amsmath,amsfonts,amssymb,rotating}

\mathcode`l="8000
\begingroup
\makeatletter
\lccode`\~=`\l
\DeclareMathSymbol{\lsb@l}{\mathalpha}{letters}{`l}
\lowercase{\gdef~{\ifnum\the\mathgroup=\m@ne \ell \else \lsb@l \fi}}%
\endgroup

\def\P{{\mathbb P}}

\newfont{\bbb}{msbm10 scaled 500}

\newfont{\bb}{msbm10 scaled 1100}

\newcommand{\xv}{{\bf x}}

\newcommand{\Xv}{{\bf X}}   

\newcommand{\Xm}{{\bf X}}

\newcommand{\Jc}{{\cal J}}

\newcommand{\Xc}{{\cal X}}

\newcommand{\js}{(j^*)}
\renewcommand{\j}{(j)}
\renewcommand{\k}{(k)}
\renewcommand{\l}{(l)}
\renewcommand{\L}{(L)}

\newcommand{\Xl}{X_{(l)}}
\newcommand{\Xk}{X_{(k)}}
\newcommand{\Xjs}{X_{(j^*)}}
\newcommand{\Xj}{X_{(j)}}

\newcommand{\Sigj}{\Sigma_{\j}^n}
\newcommand{\SigYsX}{\Sigma_{(k|j)}^n}
\newcommand{\SigYsXr}{\Sigma_{(k|j)}^n}
\newcommand{\ppj}{\rho^{(k|j)}}
\newcommand{\Uk}{U_{\k}}
\newcommand{\Ujn}{\mathbf{U}_{\k}^n}
\newcommand{\ujn}{\mathbf{u}_{\k}^n}
\newcommand{\Phin}{\boldsymbol{\Psi}_{\k}^n}
\newcommand{\Xhjn}{\widehat{\Xm}_{(k|j)}^n}
\newcommand{\Aj}{A_{(k|j)}}
\newcommand{\Bj}{B_{(k|j)}}
\newcommand{\vpi}{\lambda_{i}^{(k|j)}}
\newcommand{\vpii}{\lambda_{i}^{(k|j)}}
\newcommand{\vpip}{\lambda_{i}^{(k|j^{\star})}}

\newcommand{\Jsetk}{\mathcal{J}_{(k)}}
\newcommand{\Mk}{M_{(k)}}
\newcommand{\Rkj}{R_{(k)|(j)}}
\newcommand{\Rkjs}{R_{(k)|(j^*)}}
\newcommand{\Sk}{S_{(k)}}

\newcommand{\Un}{\mathbf{U}^n}
\newcommand{\un}{\mathbf{u}^n}

\renewcommand{\det}{{\hbox{det}}}

\newcommand{\<}{\left\langle}
\renewcommand{\>}{\right\rangle}

\def\<{\langle}
\def\>{\rangle}

\newtheorem{theorem}{Theorem}

\newtheorem{definition}[theorem]{Definition}



%% file: 0abstract.tex
%
\begin{abstract}
In  Sequential Massive Random Access (SMRA), a set of correlated sources is jointly encoded and stored on a server, and clients want to access to only a subset of the sources. Since the number of simultaneous clients can be huge, the server is only authorized to extract a bitstream from the stored data: no re-encoding can be performed before the transmission of a request.
In this paper, we investigate the SMRA performance of lossy source coding of Gaussian sources with memory. 
In practical applications such as Free Viewpoint Television, this model permits to take into account not only inter but also intra correlation between sources. 
For this model, we provide the storage and transmission rates that are achievable for SMRA under some distortion constraint, and we consider two particular examples of Gaussian sources with memory. 
\end{abstract}

%% file: 1intro.tex
%
\section{Introduction}
The amount of data available on the web is growing exponentially, as well as the number of requests to online databases (pictures, music, videos, etc.)~\cite{Hilbert_M_2011_science}.
In this context, Massive Random Access (MRA) refers to the situation where a large number of clients want to access to some content stored in a huge database. 
The MRA problem consists in finding the optimal storage requirements and transmission rates for a set of correlated sources $\{ \Xk \}_{1\leq k \leq L}$ such that the compressed sources are stored on a server, that each client requires a subset of the sources, and that this subset differs from one user to another.
For example, in Free Viewpoint Television \cite{Tanimoto_2011_ieee-spm_ftv_fvt}, the users send requests to the server in order to obtain one view within a proposed set, and they can freely switch to other views according to their fancies. 

In this paper, we consider a particular setup called Sequential Massive Random Access introduced in~\cite{Roumy_A_2015_picip_uni_lcruaci} and formally defined in~\cite{duprazArxiv17}.
SMRA has the following characteristics: (i) Sequential Access: the clients request the sources one after the other and keep their previous requests in memory, (ii) Random Access: the requests are client-dependent, (iii) Massive Access: the number of simultaneous request is huge. 
The Massive Access constraint imposes that upon request, the server cannot perform any re-encoding, but only low complexity operations such as bit extraction. 
Then, in the SMRA setup, we aim at minimizing both the storage rate of the set of sources on the server, and the transmission rates of the compressed sources transmitted from the server to the users.

SMRA is closely related to source coding with side information~\cite{sgarro77IT,Draper04,Draper07,yang10IT}, since, in the SMRA context, the previously requested source, when kept in the memory of the client, can be seen as a side information available at the decoder. 
Nevertheless, SMRA jointly optimizes the storage and transmission rates, while achievability results provided in the above works may be interpreted either in terms of storage rate~\cite{sgarro77IT,Draper07} or in terms of transmission rate~\cite{Draper04,yang10IT}. 
In~\cite{duprazArxiv17}, the joint optimization of these two rates for SMRA leads to an incremental coding scheme that achieves a double optimality. 
First, the transmission rate is equal to the rate without the Massive Access constraint, i.e. when re-encoding is allowed. Second, the storage rate is the same as without the Random Access constraint, i.e. without adaptation to the client request.

The main contribution of this paper is to provide the SMRA storage and transmission rates that are jointly achievable 
considering some distortion constraint (rates-distortion trade-off)  and realistic source models. 
In~\cite{duprazArxiv17}, lossy source coding was considered for correlated Gaussian i.i.d. sources. 
By correlated i.i.d. sources, we mean that the symbols generated by one source $\Xk$ are i.i.d. (no intra-correlation), but that the symbols generated by two sources $\Xk$ and $\Xl$ are statistically dependent (inter-correlation).
In this paper, we investigate the SMRA performance of lossy coding of Gaussian sources with both inter and intra correlation.
This problem is challenging since it requires the construction of an incremental coding scheme that leads to the double optimality (with respect to Massive Access and Random Access) while satisfying some distortion constraint for every source. 
We also consider two particular cases of Gaussian sources with memory, and we provide the achievable rates and distortions for these cases. 

The paper is organized as follows.
Section~\ref{sec:definition} introduces the SMRA framework with our source model.
Section~\ref{sec:SoA} provides the existing bounds for SMRA.
Section~\ref{sec:results} gives our main result and considers two examples.

%% file: 2definition.tex
%

\section{Lossy source coding for SMRA}\label{sec:definition}
In this section, we introduce our notations and assumptions. 
In particular, we formally define the SMRA coding scheme and we describe the considered model of Gaussian sources with memory.

\subsection{Notations}
A random source $X$ is denoted using uppercase; the source $X$ generates a sequence of random variables denoted $X_i$ using uppercase and index $i$; the realizations of the $X_i$ are denoted $x_i$ using lowercase; a random vector $\Xv$ is denoted using boldface uppercase and its realization $\xv$ is denoted using boldface lowercase. An $n$-length vector $\Xv^n=(X_1,...,X_n)$ containing elements $X_1$ to $X_n$ is denoted using superscript $n$.
The alphabet $\Xc$ of a random variable  is denoted with calligraphic letter, and with the same letter as the random variable. $|\Xc|$ denotes the cardinality of the set $\Xc$. 
In the case of multiple sources, the set $\Jc$ of source indexes is denoted using calligraphic letter. 
The $k^{th}$ source is then identified with an index inside brackets i.e. $\Xk$.

\subsection{Coding scheme definition}

\begin{figure}[t]\begin{center}
\includegraphics[width=0.6\linewidth]{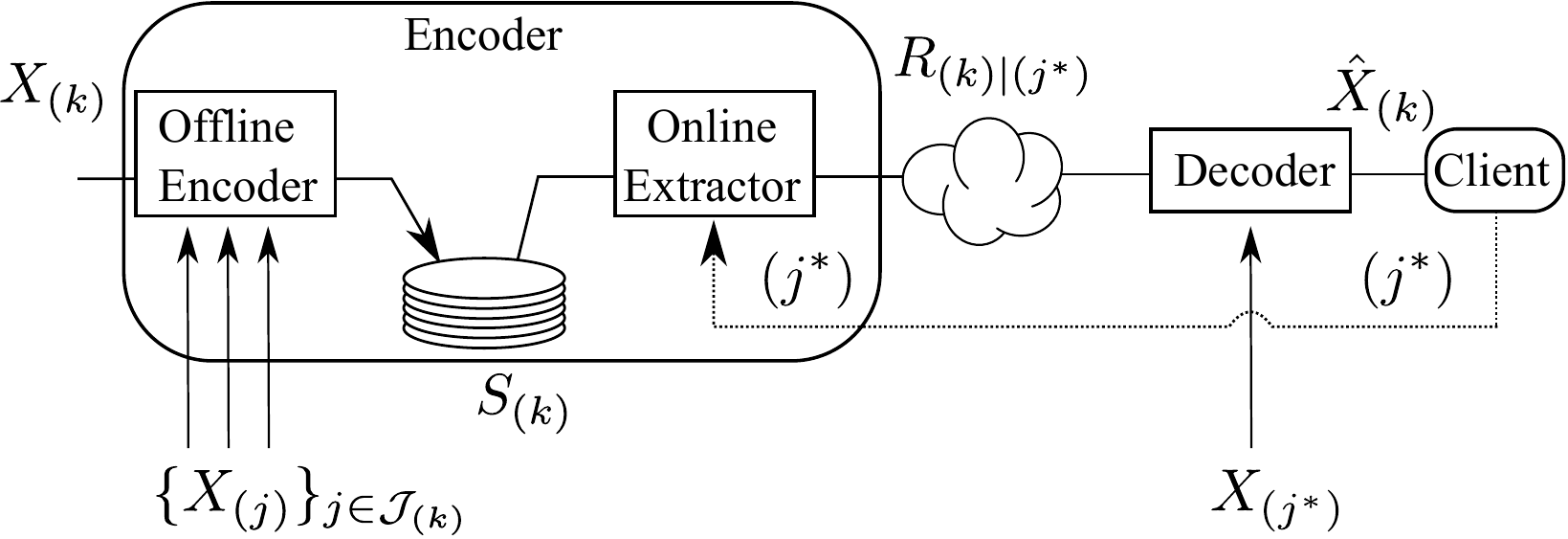}
\end{center}
\vspace*{-.5cm}
\caption{SMRA: focus on the coding of one source $\Xk$}\label{fig:MRA_problem}
\end{figure}

As initially described in \cite{duprazArxiv17}, the SMRA coding scheme is sequential in the sense that the compression of each source is performed accounting for the previously requested sources that the user will potentially have in its memory.
More formally, consider the compression of source $\Xk$ and denote by $\Jsetk$ the set that contains the indexes of the sources that can be requested just before $\Xk$. 
Here, as a first step, we consider the compression of $\Xk$ by taking into account only its potential direct predecessors and not the whole user's history of requests. The generalization to more complex sets $\Jsetk$ is left for future works.   
In practical situations, the set $\Jsetk$ depends on the constraints of the considered application.
For example, in Free-Viewpoint Television, a user may only move from one view to the neighboring left and right views, which would give $\Jsetk = \{k-1,k+1\}$. 

The overall compression scheme is depicted in Figure \ref{fig:MRA_problem} and consists of two phases.
During the first phase, the source $\Xk$ is encoded by the offline encoder into an incremental index sequence at storage rate $\Sk$ bits/source symbol under the assumptions that: (i)  all the realizations of the sources in $\Jsetk$ are known, (ii) only one source with index in $\Jsetk$ will be available at the client's decoder. 
The second phase starts when the client requests $\Xk$ and specifies the index $j^* \in \Jsetk$ of the source stored in its memory. 
The server then extracts an index subsequence at rate $\Rkjs$ (online extractor) and sends it over a noiseless link to the user. 
To finish, the decoder estimates the requested source from the received index subsequence and from the available source $\Xjs$.  

The SMRA code construction is incremental because the server will have to adapt the rate (by extracting a subsequence of indices) to any possible previous request $j^{\star} \in \Jsetk$, without re-encoding. 
We now formally define the SMRA code and the jointly achievable storage rates, transmission rates, and distortions for this code.
\begin{definition}[SMRA code] 
\label{def:SMRAcode}
A $( (2^{n\Sk}, ( 2^{n \Rkjs} )_{j^{\star} \in \Jsetk } )_{1 \le k \le L},n)$ \emph{SMRA code} for the set of discrete general sources $\{ \Xk \}_{1\leq k \leq L}$  consists, for each source $\Xk$, $1 \leq k \leq L$, of
\begin{itemize}
\item an \emph{offline encoder} $h^{\text{\emph{off}}}_{\k}$ that assigns a sequence of $\Mk=|\Jsetk|$ indices 
to the set of vectors $\left(\xv^n_{\k}, ( \xv^n_{\js}  )_{j^{\star} \in \Jsetk }\right) \in \Xc^n \times \Xc^{n\times\Mk}$ 
\begin{subequations}\label{eq:enc-off}
\begin{align}
h^{\text{\emph{off}}}_{\k}: \Xc^n \times \Xc^{n \times \Mk} &\to  \prod_{m=1}^{\Mk} \{ 1, \ldots, 2^{nr_{k,m}} \} \\
\xv^n_{\k}, ( \xv^n_{\js}  )_{j^{\star} \in \Jsetk } & \mapsto ( i_1, \ldots , i_{\Mk} ) 
\end{align}
\end{subequations} 
where $\Sk = r_{k,1} + r_{k,2} + \ldots + r_{k,\Mk}  $.
\item a set of $\Mk$ \emph{online extractors} $h^{\text{\emph{on}}}_{{(k)|\js}}$, $j^{\star} \in \Jsetk$, that extract a subsequence of indices from the sequence of indices $( i_1, \ldots , i_{\Mk} )$
\begin{subequations}\label{eq:enc-on}
\begin{align}
h^{\text{\emph{on}}}_{{(k)|\js}}: \prod_{m=1}^{\Mk} \{ 1, \ldots, 2^{nr_{k,m}}\}&\to  \prod_{m \in \mathcal{I}_{(k)|\js} } \{ 1, \ldots, 2^{nr_{k,m}} \} \\
 ( i_1, \ldots , i_{\Mk} ) & \mapsto ( i_m )_{m \in \mathcal{I}_{(k)|\js}}
\end{align}
\end{subequations} 
where $\mathcal{I}_{(k)|\js} \subseteq \{1, \cdots, \Mk\}$, and $\Rkjs = \sum_{m \in \mathcal{I}_{(k)|\js}} r_{k,m}  \le \Sk$
\item a set of $\Mk$ \emph{decoders} $g_{{(k)|\js}}$, $j^{\star} \in \Jsetk$, that, given the source realization $\xv^n_{\js} $, assign an estimate $\hat \xv^n_{\k|\js}$ to each received subsequence of indices
\begin{subequations}\label{eq:dec}
\begin{align}
g_{{(k)|\js}}:    \prod_{m \in \mathcal{I}_{(k)|\js} } \{ 1, \ldots, 2^{nr_{k,m}} \}    \times \Xc^n &\to \Xc^n \\
 ( i_m )_{m \in \mathcal{I}_{(k)|\js}} ,  \xv^n_{\js} & \mapsto \hat \xv^n_{\k|\js}
\end{align}
\end{subequations} 
\end{itemize}
\end{definition}

\begin{definition}[Rates-distortion region for SMRA code] 
\label{def:SMRA_achievablerate}
Consider a distortion measure $d: \Xc \times \Xc \rightarrow \mathbb{R}$.
The tuple
$
 \left((\Sk,( \Rkjs )_{j^{\star} \in \Jsetk }, ( D_{(k)|\js} )_{j^{\star} \in \Jsetk })_{1\le k\le L}\right)
$
is said to be \emph{achievable for SMRA} if there exists a sequence of SMRA codes such that 
\begin{equation}
\forall k\in \{1,\cdots,L\}, \forall j^{\star} \in \Jsetk, \lim_{n \rightarrow \infty} E\left[ \frac{1}{n} \sum_{i=1}^n d \left( X_{\k,i} ,\hat{X}_{\k|\js,i}  \right)  \right] \leq D_{(k)|\js},
\end{equation}
where the expectation is taken over $(\Xv_{\k}^n, \Xv_{\js}^n)$. 

\end{definition}

The main particularity of the SMRA code definition resides in the combination of two encoding mappings: a standard offline encoder that produces the sequence of coded indexes, and a novel online extractor that can only extract a part of the coded indexes. 
The online extractor is a very simple operation introduced because re-encoding is not desirable in massive access to data.
The above definition suggests that the encoder and the extractor should be jointly designed in order to minimize both storage $\Sk$ and transmission $\Rkjs$  rates involved in the definition.

\subsection{Gaussian source model with memory}\label{sec:gaussian_model}
In this paper, we derive the tuple of jointly achievable rates $\Sk$, $\Rkjs$, and distortions $D_{(k)|\js}$ for a Gaussian source model with memory which we now describe.
For all $j \in \Jsetk $, we assume that the source $\Xj$ generates Gaussian vectors of length $n$ as $\Xv^n_{\j} \sim \mathcal{N}(0,\Sigj)$. The covariance matrices $\Sigj$ are positive-definite.
For all $j \in \Jsetk $, the statistical dependence between $\Xj$ and $\Xk$ is described by $(\Xv_{\k}^n|\Xv_{\j}^n = \xv^n_{\j}) \sim \mathcal{N}(\xv^n_{\j},\SigYsX)$.
The covariance matrices $\SigYsX$ are assumed to be positive-definite Toeplitz matrices with expression
\begin{equation}\label{eq:cov_matrix}
 \SigYsX = \begin{bmatrix}
                \ppj_1 & \ppj_2 & \ppj_3 & \cdots & \ppj_{n} \\
                \ppj_2 & \ppj_1 & \ppj_2 & \cdots & \ppj_{n-1} \\
                \cdots & \cdots & \cdots & \cdots & \cdots \\
                \ppj_{n} & \ppj_{n-1} & \cdots & \ppj_2 & \ppj_1
               \end{bmatrix} .
\end{equation}
Let $\vpi$, $\ell = 1,\cdots, n$ be the $n$ eigenvalues of $\SigYsX$. 
For Gaussian sources, $\mathcal{X} = \mathbb{R}$ and we consider the quadratic distortion measure defined $\forall (x,y) \in \mathbb{R}\times \mathbb{R}$ by $d(x,y) = (x-y)^2$. 
This model captures the dependencies between components for a wide range of sources with stationarity in the memory.
Finite memory can be considered by setting $\ppj_{i} = 0$ for all $i$ greater than a given integer. 
For this model,~\cite{berger71b} provided the rate-distortion region in the standard case of lossy source coding without side information.

The sources $\Xj$ that may serve as side information for $\Xk$ were also reconstructed with a certain distortion.
The set $\Jsetk$ then contains all the possible distorded versions of the sources that can be available at the decoder when $\Xk$ is requested. 
With the model introduced in this section, we assume a Gaussian model between the source $\Xk$ and the sources with distortion contained in $\Jsetk$.
The expressions of the covariance matrices $\SigYsX$ given in~\eqref{eq:cov_matrix} hence depend on the distortion levels in these sources. 
In the following, we first describe already existing information-theoretic results for particular cases of this model, we then provide our main result of achievability for SMRA. 

%% file: 3stateofart.tex
%
\section{Source Coding performance: knowns bounds}\label{sec:SoA}

The main particularity of the SMRA coding scheme (see Figure \ref{fig:MRA_problem}) lies in the splitting of the encoder into two parts. 
The offline encoder has access to all the data but does not know the index $j^\star$ of the source available at the decoder, and the online encoder has access to the coded sequence of indexes and to the index $j^\star$. 
The storage and transmission rates that are achievable for SMRA have been derived in \cite{duprazArxiv17}. 

For instance, let us consider a set of $L$  i.i.d. sources such that the joint distribution can be factorized as
$P(\xv^n_{(1)},...,\xv^n_{\l},...,\xv^n_{\L}) = \prod_{i=1}^n P(x_{(1),i},...,x_{\L,i})$.
When a client requests the source $\Xk$ and indicates that the source $\Xjs$ is available at its decoder, we show that~\cite{duprazArxiv17} 
\begin{subequations}\label{eq:RS_noniid}
\begin{align}
\Sk&\ge \displaystyle \max_{j \in \Jsetk} H \left(  \Xk | \Xj \right) \\
\Rkjs &\ge H \left(  \Xk | \Xjs \right),
\end{align}
\end{subequations}
where $H(\Xk | \Xjs)$  is the conditional entropy of the source $\Xk$ given $\Xj$.
As a result, the transmission rate $H \left(  \Xk | \Xjs \right)$ is the same as if re-encoding was allowed (optimality despite the Massive Access constraint). 
Moreover, this optimal transmission rate can be achieved while keeping the storage rate at its lowest possible value $\max_{j \in \Jsetk} H \left(  \Xk | \Xj \right)$, which is much smaller than the rate $\sum_{j \in \Jsetk} H (  \Xk | \Xj )$ that is required when storing a different codeword for all possible pairs $(\Xj,\Xk)_{j\in \Jsetk}$ (optimality despite the Random Access constraint). A similar result holds for lossy compression of i.i.d. sources.

The above cases consider sources with no intra correlation (the source components $X_{\k,i}$ are i.i.d.) but with inter correlation ($X_{\k,i}$ and $X_{\j,i}$ are statistically dependent). 
However, real data such as videos always contain intra correlation and it is of great interest to see if the optimality of SRMA remains in this case.
Consider the Gaussian model with intra correlation described in Section~\ref{sec:gaussian_model} and assume that $\Xjs$ is the previously requested source. 
For this model, the marginal Karhunen Loeve Transform (KLT)~\cite{gastpar06IT} derived from the covariance matrix of $\Xv_{\k}^n$ will not take into account the inter correlation of $\Xk$ and $\Xjs$ and it will lead to a suboptimal transmission rate.   
On the other hand, the conditional KLT~\cite{gastpar06IT} derived from the covariance matrix $\Sigma_{(k|j^{\star})}$ would lead to an optimal transmission rate. 
However, the conditional KLT cannot be applied in the SMRA coding scheme, since the offline encoder does not know the index $j^{\star}$ of the source that will be available at the decoder. 
In the remaining of the paper, we study the SMRA coding of the Gaussian sources described in Section~\ref{sec:gaussian_model}, and we propose an incremental coding scheme that applies to sources with inter and intra correlation.


%% file: 4results.tex
\section{Lossy Source Coding for Correlated Gaussian Vectors}\label{sec:results}
%

The following theorem states our main result by providing  the achievable tuple of rates and distortions for non i.i.d. Gaussian sources for SMRA. 
The proof is given in Section~\ref{sec:proof_th}.
\begin{theorem}\label{th:rate_dist_region}
For given parameters $\delta_k$, $\theta_k$ $(k \in \{ 1,\cdots,L \})$,
the  rates-distortions tuple $\left((\Sk,( \Rkjs, D_{(k)|(j^*)} )_{j^* })_{k}\right)$ is achievable for Gaussian sources for SMRA if $\forall k \in \{ 1,\cdots,L \}, \forall j^{*} \in \Jsetk$,
\begin{align}\label{eq:result_th}
\Rkjs(\delta_k,\theta_k) &\geq \lim_{n \rightarrow \infty} \frac{1}{n} \sum_{i=1}^n \max\left(0,\frac{1}{2} \log_2 \frac{\vpip}{\theta_k}\right)   \\
\Sk(\delta_k,\theta_k) &\geq \lim_{n \rightarrow \infty} \frac{1}{n} \sum_{i=1}^n \max\left(0, \max_{j \in \Jsetk} \frac{1}{2} \log_2 \frac{\vpi}{\theta_k}\right) \\ 
 D_{(k)|(j^*)}(\delta_k,\theta_k)  & \leq \lim_{n \rightarrow \infty} \frac{1}{n} \sum_{\ell=1}^n \min \left(\theta, \frac{\vpip \delta_k}{\vpip + \delta_k} \right) .
\label{eq:RSWZ}
\end{align}
given that the limits exist.
\end{theorem}
In the above theorem, we notice that the transmission rate $\Rkjs(\delta_k,\theta_k)$ corresponds to the Wyner-Ziv rate-distortion function for a given target distortion $D_{(k)|(j^*)}(\delta_k,\theta_k)$ when $\Xjs$ is the only possible side information.
The storage rate $\Sk(\delta_k,\theta_k)$ is given by the mean of the worst possible rates $ \frac{1}{2} \log_2 \frac{\vpi}{\theta_k}$ for each components $i\in\{1,\cdots,n\}$. 

In Theorem~\ref{th:rate_dist_region}, the parameter $\theta_k$ comes from the waterfilling problem of allocating the rate between source components in order to achieve a distortion constraint in expectation. 
The parameter $\delta_k$ is the distortion of an individual component when no previous request is available at the decoder. 
When a previous request $\Xjs$ is available at the decoder, the source $\Xk$ can be reconstructed with a distortion $ D_{(k)|(j^*)} \leq \delta_k$ that depends on the parameter $\delta_k$ and on the statistics between $\Xk$ and $\Xjs$.
It is worth noting that all the distortion levels  $D_{(k)|(j)}$ only depend on the eigenvalues of $\SigYsX$ and on the unique parameter $\delta_k$. 
In particular, it is not possible to achieve a particular distortion for a given $\Xjs$ without affecting all the distortions for the other possible $\Xj$.
This is due to the incremental aspect of SMRA, as can be seen in the following proof.


\subsection{Proof of achievability}\label{sec:proof_th}

 \vspace{-0.3cm}

\paragraph{Test-channel:}
We consider the following test channels
\begin{align}\label{eq:modelU}
 \Ujn & = \Xv_{\k}^n + \Phin \\ \label{eq:modelXh}
 \forall j \in \Jsetk, ~ \Xhjn & = \Aj \Ujn + \Bj \Xv_{\j}^n
\end{align}
where  $\Phin \sim \mathcal{N}(0,\delta_k I_n)$, $I_n$ is the identity matrix of size $n\times n$. $\Aj$ and $\Bj$ are $n\times n$ matrices such that 
\begin{align}
 \Aj & = \SigYsXr (\delta_k I_n + \SigYsXr)^{-1} \\
 \Bj & = (I_n + \delta_k \SigYsXr)^{-1} \SigYsXr (\Sigj)^{-1}.
\end{align}
For all $i \in \{1,\cdots, n \}$ and $j\in \Jsetk $, this test channel gives individual distortions
\begin{equation}\label{eq:indiv_dist}
E\left[ (X_{\k,i} - \hat{X}_{(k|j),i})^2 \right] =  \left( \left(\frac{1}{\delta_k} I_n + (\SigYsXr)^{-1}\right)^{-1}\right)_{i,i} = \frac{\vpii \delta_k}{\vpii + \delta_k} .
\end{equation}

 \vspace{-0.3cm}

\paragraph{Random code generation:}
Generate $2^{nr_0}$ sequences $\Ujn$ at random according to~\eqref{eq:modelU}.
The distribution of $\Ujn$ does not depend on the possible previous requests $\Xj$.
Denote by $\mathcal{C}$ the set of generated sequences $\ujn$ and index them with $s \in \{ 1,\cdots,2^{nr_{0}}\}$. 
Assign each $\ujn(s) \in \mathcal{C}$ to $\Mk$ incremental bins, following the same process as in the proof of~\cite[Theorem 6]{duprazArxiv17}.
In order to construct the incremental bins, consider the source reordering function $\pi: \Jsetk \rightarrow \{1,\cdots, \Mk\}$, $j \rightarrow \pi(j) $. 
We denote $m = \pi(j)$, and the reordering function $\pi$ is such that $\bar{I}(\mathbf{X}_{\pi^{-1}(m)};\mathbf{U}_{\k}) \leq \bar{I}(\mathbf{X}_{\pi^{-1}(m-1)};\mathbf{U}_{\k}), ~ \forall m\in\{2,\cdots \Mk\}, $ where $\bar{I}(.;.)$ is the spectral mutual information defined in~\cite[Section 5.4]{han2003b}.
The size of the $\Mk$ incremental bins is defined by values $r_{m}$ such that at the $m$-th level, there are $2^{n(r_1 + \cdots + r_{m})}$ bins. 
This defines $\Mk$ mappings $f_{(k|j)}(\un_{\k}) = (i_1,\cdots,i_{\pi(j)})$, $j \in \{1,\cdots, \Mk\}$ where the $(i_1,\cdots,i_{\pi(j)})$ are the indices of the successive bins to which $\ujn$ belongs.

 \vspace{-0.3cm}

\paragraph{Encoding:}
Given a sequence $\xv_{\k}^n$, find a sequence $\ujn(s) \in \mathcal{C}$ such that $(\xv_{\k}^n, \ujn(s)) \in T_{\varepsilon,n}^{(1)}(\Xk,\Uk)$, where
 \begin{equation}\label{eq:Adef}
  T_{\varepsilon,n}^{(1)}(\Xk,\Uk) = \left\{ (\xv_{\k}^{n},\ujn) \left| \frac{1}{n} \log \frac{P(\ujn|\mathbf{x}_{\k}^{n})}{P(\ujn)} < r_0 - \varepsilon \right. \right\} .
 \end{equation}
The offline encoder then sends to the storage unit the index sequence $(i_1,\cdots,i_{\Mk}) $ obtained for $\ujn(s)$.
Upon request of the source $X$ and previous request $j$, the online extractor sends to the user the index sequence $ (i_1,\cdots,i_{\pi(j)})$ for $\ujn(s)$.

 \vspace{-0.3cm}

\paragraph{Decoding:}
Given the received index sequence $(i_{1},...,i_{\pi(j)})$ and the side information $\xv^n_{\j}$, declare $\hat{\mathbf{u}}_{\k}^n = \ujn(s)$ if there is a unique pair of sequences $(\xv^n_{\j},\ujn(s))$ such that $f_{{(k|j)}}(\ujn(s)) =(i_{1},...,i_{\pi(j)})$ and  $(\xv^n_{\j},\un_{\k}(s)) \in T_{\varepsilon,n}^{(2)}(X_{\j},\Uk)$ where
 \begin{equation}\label{eq:Adef2}
  T_{\varepsilon,n}^{(2)}(X_{\j},\Uk) = \left\{ (\xv_{\j}^{n},\ujn) \left| \frac{1}{n} \log \frac{P(\ujn|\xv_{\j}^{n})}{P(\ujn)} < \sum_{i=1}^j r_i - \varepsilon \right. \right\} .
 \end{equation}
Then compute $\hat{\mathbf{x}}_{(k|j)}^n$ from $\hat{\mathbf{u}}_{\k}^n$ and $\xv^n_{\j}$ according to~\eqref{eq:modelXh}.

 \vspace{-0.3cm}

\paragraph{Probability of error:}
We define the error events: 
\begin{align}\notag 
& E_{0,1}= \{ \nexists s \mbox{ such that } (\Xv_{\k}^n,\ujn(s)) \in  T_{\varepsilon,n}^{(1)}(\Xk,\Uk) \} \\ \notag 
& E_{0,2}= \{ (\Xv_{\k}^n,\ujn(s)) \in  T_{\varepsilon,n}^{(1)}(\Xk,\Uk) \mbox{ but } (\Xv^n_{\j},\ujn(s)) \notin  T_{\varepsilon,n}^{(2)}(X_{\j},\Uk)\} \\ \notag 
& E_j= \{ \exists s' \neq s: f_{(k|j)} (\ujn(s')) =  f_{(k|j)} (\ujn(s))   \mbox{ and } (\Xv^n_{\j},\ujn(s')) \in T_{\varepsilon,n}^{(2)}(X_{\j},\Uk)\}, \ \ \ \forall j \in \Jc
\end{align}
By the same derivation as in the proof of~\cite[Theorem 6]{duprazArxiv17}, we show that $\P(E_j) \rightarrow 0$ as $n \rightarrow \infty$, $\forall j \in \Jc$. 
By the definitions of the spectral mutual information $\bar{I}(\mathbf{X}_{\k};\mathbf{U}_{\k})$, see~\cite[Section 5.4]{han2003b}, and of the set $T_{\varepsilon,n}^{(1)}$ in~\eqref{eq:Adef} we show that if $r_0 \geq \bar{I}(\mathbf{X}_{\k};\mathbf{U}_{\k})$, then $\P(E_{0,1}) \rightarrow 0$ as $n \rightarrow \infty$.
With the same arguments and from the definition of $ T_{\varepsilon,n}^{(2)}$ in~\eqref{eq:Adef2}, we show that $\P(E_{0,2}) \rightarrow 0$ as $n \rightarrow \infty$ if $\sum_{i=1}^j r_i \geq   \bar{I}(\mathbf{X}_{\k};\mathbf{U}_{\k})  - \bar{I}(\mathbf{X}_{\j};\mathbf{U}_{\k})$.
At the end and from the two above rate conditions, the decoding error probability $P_{error}^n = \P ( E_{0,1} \bigcup E_{0,2} \bigcup  \cup_{j\in \Jc}  E_j)\to 0$ as $n\to\infty$.

\paragraph{Distortion and rate computation:}
First, from the individual distortions~\eqref{eq:indiv_dist} and from the error probability analysis, the overall distortion for the sequence $\Xv_{\k}^n$  can be calculated for all $j\in\Jsetk$ as
\begin{equation}\label{eq:finalD}
\frac{1}{n} \sum_{i=1}^n E[(X_{\k,i} - \hat{X}_{(k|j),i})^2 ] \leq (1- P_{error}^n)\frac{1}{n}\sum_{i=1}^n \frac{\vpii \delta_k}{\vpii + \delta_k}  + P_{error}^n \delta_{max}
\end{equation}
where $\delta_{max}$ is a constant that represents the maximum possible distortion over a given component, and the expectation is calculated given that $\Xj$ is available at the decoder.
Then, from the definition of the spectral mutual information in~\cite[Section 5.4]{han2003b} and by the ergodicity of the considered Gaussian sources, $\bar{I}(\mathbf{X}_{\k};\mathbf{U}_{\k})  - \bar{I}(\mathbf{X}_{\j};\mathbf{U}_{\k}) = \lim_{n \rightarrow \infty} \frac{1}{n}h(\Un_{\k}|\Xv_{\k}^n) - \frac{1}{n}h(\Un_{\k}|\Xv_{\j}^n)$.
From~\cite{gastpar06IT}, we can then express
\begin{equation}\label{eq:finalR}
 \frac{1}{n}h(\Un_{\k}|\Xv_{\k}^n) - \frac{1}{n}h(\Un_{\k}|\Xv_{\j}^n) = \frac{n}{2} \log_2 \frac{\det (\SigYsXr + \delta_k I_n)}{\det(\delta_k I_n)} = \frac{1}{n} \sum_{i=1}^n \frac{1}{2} \log_2 \frac{\vpii}{d_{i}},
\end{equation}
where $\det(.)$ is the determinant of the matrix in argument, and $d_{i} = \frac{\vpii\delta_k}{\vpii + \delta_k}$. 
At the end, taking the limits when $n\rightarrow \infty$ in~\eqref{eq:finalD} and~\eqref{eq:finalR}, and expressing the rate-allocation optimization between the individual components $X_{\k,i}$ of $\Xv_{\k}$ gives the rate and distortion expressions in~\eqref{eq:result_th}.
It can be seen from~\eqref{eq:finalD} and~\eqref{eq:finalR} that the rate and the distortion are allocated component by component. 
This operation leads to the expression of $\Sk$ in~\eqref{eq:result_th} in which the maximum over the $j \in \Jsetk$ is taken component by component.

%

\subsection{Examples}

\begin{figure*}[t]
\begin{center}
  \subfigure[~]{ \includegraphics[width=0.45\linewidth]{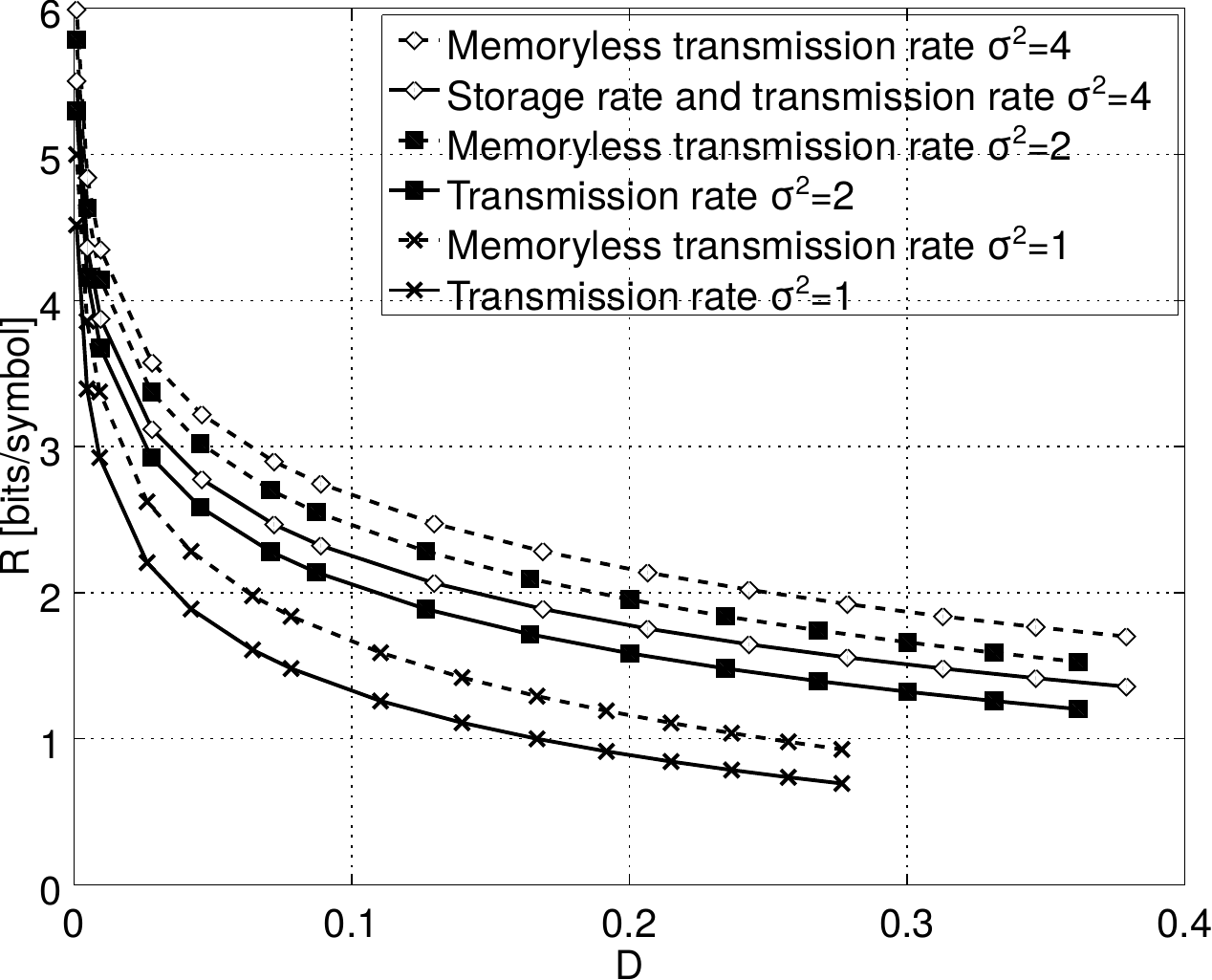}}
  \subfigure[~]{ \includegraphics[width=0.45\linewidth]{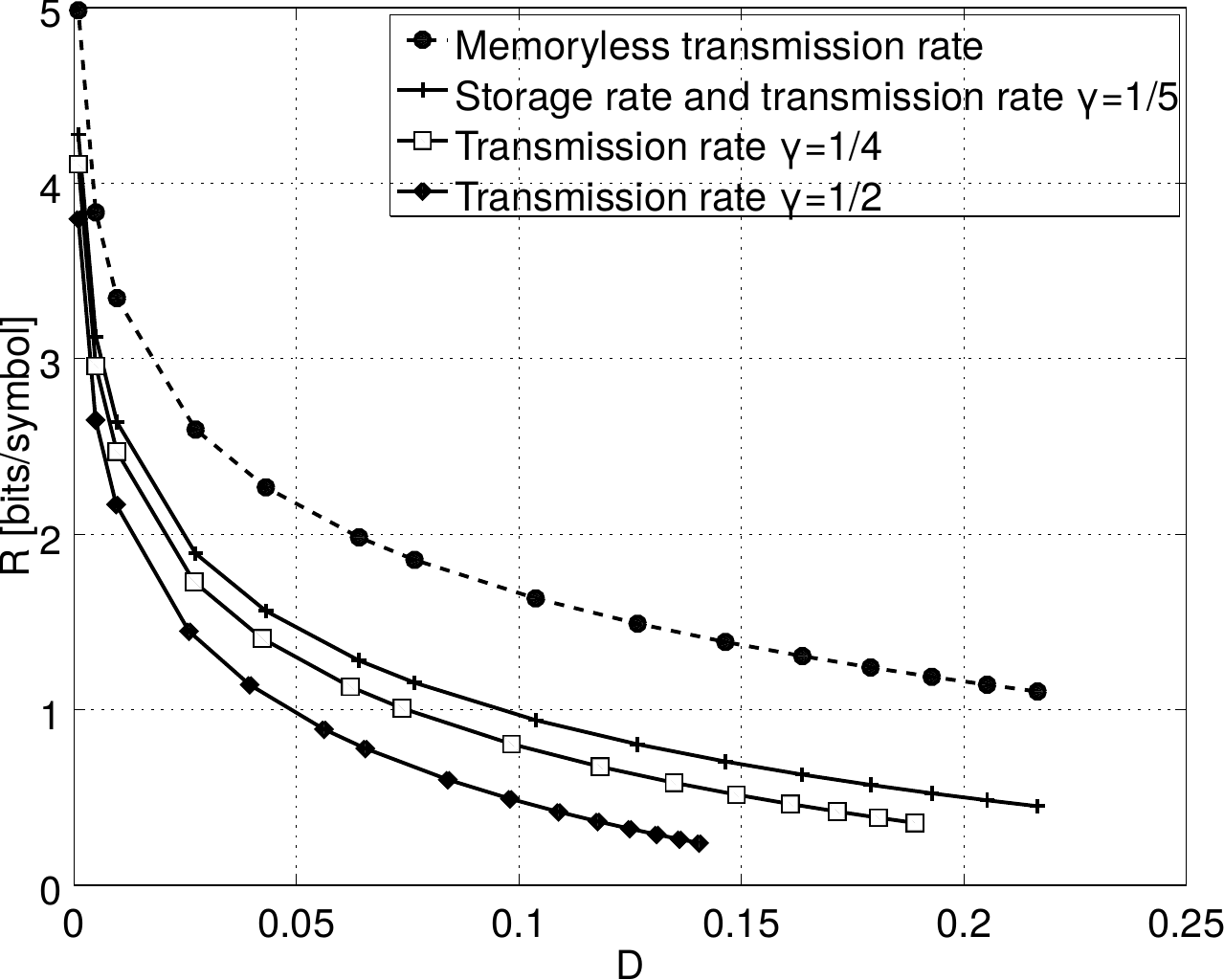}}
\end{center}
\caption{Rates $\Sk$ and $\Rkj$ with respect to distortions $D_{\k|\j}$ for three possible previous requests $\Xj$: (a) Nearest  Neighbor correlation model, (b) First-order Markov model }
\label{fig:allfigs}
\end{figure*}

\vspace{-0.3cm}

As an example, we consider one source $\Xk$ and $M=3$ possible previous requests $\Xj$. 
We consider two Gaussian models that are particular cases of our model introduced in Section~\ref{sec:gaussian_model}. 
The two considered cases have been introduced in~\cite{berger71b} for the standard case of lossy source coding without side information.  
In order to obtain approximations of the rate and distortion expressions provided in Theorem~\ref{th:rate_dist_region}, we computed the eigenvalues of the three Toeplitz matrices obtained with each considered model at length $N=1000$. 

\vspace{-0.3cm}

\paragraph{Nearest Neighbor correlation}
We first consider the Gaussian model of Section~\ref{sec:gaussian_model} with $\ppj_1 = \sigma_j^2$, $\ppj_2 = \sigma_j^2/2$, and $\ppj_i = 0$, $\forall i>2$.  
This model assumes that each component $i$ in $\Xv_{\k}$ is only correlated with the components $i-1$, $i$, and $i+1$ of $\Xv_{\j}$. 
In order to completely define the model for each of the three possible previous requests, we set $\sigma_1^2 = 1$,  $\sigma_2^2 = 2$,  $\sigma_3^2 = 4$.
The rate-distortion functions for SMRA for this model are represented in Figure~\ref{fig:allfigs} (a). 
As expected, the storage rate is superimposed with the worst possible transmission rate. 
Figure~\ref{fig:allfigs} (a) also shows the rate-distortion functions for SMRA for the memoryless Gaussian model ($\ppj_1 = \sigma_j^2$  and $\ppj_i = 0$, $\forall i>1$), which illustrates the gain at taking the memory into account. 

\vspace{-0.3cm}

\paragraph{First-order Markov source}
We now consider the Gaussian model with $\ppj_i = \sigma^2 |\gamma_j|^{i-1}$, with $-1<\gamma_j<1$, for all $i\in \{1,\cdots, n\}$. 
This model assumes that all the components of  $\Xv_{\k}$ and $\Xv_{\j}$ are correlated with a level $\sigma^2  |\gamma_j|^i$ decreasing with $i$.  
For this model, we set $\sigma^2 = 1$ $\gamma_1 = 1/2$, $\gamma_2=1/4$, and $\gamma_3=1/5$. 
The corresponding rate-distortion functions for SMRA are represented in Figure~\ref{fig:allfigs} (b), as well as the rate-distortion functions for the memoryless case.
In this case, we observe that the memoryless rate-distortion functions are the same whatever the previously requested source $\Xj$ available at the decoder, while taking the memory into account permits a decrease in the transmission rates and distortions. 

%% file: 5conclusion.tex
%

\section{Conclusion}
In this paper, we considered SMRA source coding for Gaussian sources with memory, and we provided the achievable storage and transmission rates for this problem. 
For this source model, the transmission rate is equal to the rate without the Massive Access constraint, and the storage rate is equal to the rate without the Random Access constraint, as for the lossless i.i.d. case. 
However, with the incremental scheme considered in the paper, it is not possible to achieve a particular distortion for a given previous request without affecting all the distortions for the other possible previous requests. 
This issue will be tackled in future works.

\section*{Acknowledgement}
This work has received a French government support granted to the Cominlabs excellence laboratory and managed by the National Research Agency in the ``Investing for the Future'' program under reference ANR-10-LABX-07-01.